\documentstyle[aps,epsf,preprint,graphicx,prl]{revtex}

\begin{document} 
\draft

\def\beq{\begin{equation}}
\def\eeq{\end{equation}}
\def\beqn{\begin{eqnarray}}
\def\eeqn{\end{eqnarray}}
\def\btimes {\mbox{\boldmath $\times$}}
\def\bbox {\mbox{\boldmath $\box$}}
\def\bvarphi {\mbox{\boldmath $\varphi$}}
\def\ed{\end{document}}

\def\veps {{\varepsilon}}
\def\I {{\bf I}}
\def\II {{\bf II}}
\def\III {{\bf III}}
\def\IV {{\bf IV}}
\def\V {{\bf V}}
\def\VI {{\bf VI}}
\def\J {{\bf J}}
\def\H {{\bf H}}
\def\E {{\bf E}}
\def\1 {{\bf 1}}
\def\2 {{\bf 2}}
\def\3 {{\bf 3}}
\def\P {{\bf P}}
\def\r {{\bf r}}
\def\k {{\bf k}}
\def\p {{\bf p}}
\def\n {{\bf n}}
\def\A {{\bf A}}
\def\bv {{\bf v}} 
\def\AAN {$\!\!\!$ A$^{^{\!\!\!\!\! {\tiny {\circ}}}}$}
\def\aaN {$\!\!$ a$^{^{\!\!\!\! {\tiny {\circ}}}}$}

\title{Pathway from condensation via fragmentation 
to fermionization of cold bosonic systems}

\author{Ofir E. Alon\footnote{E-mail: ofir@tc.pci.uni-heidelberg.de}
 and Lorenz S. Cederbaum\footnote{E-mail: Lorenz.Cederbaum@urz.uni-heidelberg.de}}
\address{Theoretische Chemie, Physikalisch-Chemisches Institut, Universit\"at Heidelberg,\\
Im Neuenheimer Feld 229, D-69120 Heidelberg, Germany}

\maketitle

\begin{abstract}
For small scattering lengths, 
cold bosonic atoms form a condensate the density profile of which is smooth.
With increasing scattering length, the density {\it gradually} acquires more and more oscillations.
Finally, the number of oscillations equals the number of bosons and the system becomes {\it fermionized}.
On this pathway from condensation to fermionization intriguing phenomena occur, 
depending on the shape of the trap.
These include macroscopic fragmentation and
 {\it coexistence} of condensed and fermionized parts that are separated in space.
\end{abstract}

The experimental observations of Bose-Einstein condensates (BECs) \cite{Ketterle,Wieman} 
have promoted a contemporary search for the physics governing {\it trapped} cold bosonic systems \cite{Stringari_book}.
For weakly-interacting bosons, Gross-Pitaevskii theory \cite{Gross_Pitaevskii} provides an excellent description of the ground state.
For the other extreme case, namely, for impenetrable bosons in one dimension (1D) -- so-called Tonks-Girardeau gas -- 
the mapping between bosons and free (spinless) fermions is exact \cite{Girardeau_1960}
and provides the ground state.
Clearly, the Gross-Pitaevskii theory must fail in 1D somewhere in between 
these extreme cases, also see \cite{Kolomeisky,Petrov}.
The crossover from a BEC to the Tonks-Gerardeau regime has been a subject of considerable interest
and its study has mainly focused on harmonic traps 
 \cite{Petrov,Olshanii_2001,Girardeau_Wright_2001,Blume_2002,Gangardt_2003,Blume_2004_PRL,Brand_2004}.
With the recent realization of a Tonks-Girardeau gas \cite{IB_nature},
it is anticipated that theoretical predictions would be captured in the lab. 

What happens on the pathway from condensation to fermionization
and how, at all, to follow this pathway for a {\it general trap potential} 
have yet to be addressed.
In fact, even for harmonic traps oscillations of the density due to ``fermionization'' of
the bosonic system have not been recovered for finite scattering lengths.
This Letter addresses the crossover from condensation to fermionization in a general trap potential.
We will show that 
(a) the density {\it gradually} acquires oscillations as the scattering length increases,
the number of which equals the number of bosons when the system becomes {\it fermionized} (enters the 
Tonks-Girardeau regime);
(b) the pathway from condensation to fermionization depends on the shape of the potential.
In a general trap, intriguing phenomena not encountered in the homogeneous or harmonic system can occur.
For instance, in asymmetric double-well potentials a cold bosonic system 
can be condensed (coherent) in one well and fermionized in the second one.

We would like to follow the pathway from weakly- to strongly-interacting bosons 
in real space and obtain a wavefunction picture
of the ground state of cold atoms in the trap.
This will allow us to monitor directly the {\it fine} changes
in the spatial density of interacting bosons as the inter-particle 
interacting increases.
Gross-Pitaevskii theory, which is an excellent mean-field for weakly-interacting boson systems, 
fails to describe the system as the interaction increases \cite{Kolomeisky,Petrov}.
There is, however, a more general mean-field for bosons
that employs a single configurational wavefunction (also termed permanent) \cite{LA_PLA}.
Generally, we may put $n_1$ bosons to reside in one orbital, $\phi_1(\r)$,
$n_2$ bosons to reside in a second orbital, $\phi_2(\r)$, and so on,
distributing the $N$ atoms among $1 \le n_{orb} \le N$ orthogonal orbitals.
We are going to look for the 
ground state of the $N$-boson system in the subspace of {\it all} 
possible configurations:
$\Psi = \hat {\cal S} \phi_1(\r_1) \cdots \phi_1(\r_{n_1})
                      \phi_2(\r_{n_1+1}) \cdots \phi_2(\r_{n_1+n_2}) \cdots
                      \phi_{n_{orb}}(\r_N)$,
where 
$\hat {\cal S}$ is the symmetrization operator.
Gross-Pitaevskii theory -- where all bosons reside in a single orbital --
is now only one possible configuration in which the system might be found.

Let the many-body Hamiltonian describing $N$ bosons in a general trap potential be
$\hat H =  \sum_{i=1}^{N} \left[ \hat T(\r_i) +  V(\r_i) \right] +
 \sum_{i>j=1}^N U(\r_i-\r_j)$.
Here, $\r_i$ is the coordinate of the $i$-th particle,
$\hat T(\r_i)$ and $V(\r_i)$ stand for the kinetic energy and trap potential, respectively,
and  $U(\r_i-\r_j) = g \delta(\r_i-\r_j)$ 
describes the pairwise contact interaction between the $i$-th and $j$-th atoms,
where $g$ is proportional to the s-wave scattering length.
Taking the expectation value of the Hamiltonian with respect to $\Psi$
leads to the following energy functional \cite{LA_PLA}:
\beqn\label{energy_functional}
 E_N[n_{orb},\{n_i\},\{\phi_i\}] &=& \sum_i^{n_{orb}} 
\left\{n_i  \int \phi_i^\ast(\r)  \left[ \hat T(\r_i) +  V(\r_i) \right]  \phi_i(\r) d\r  
+ g \frac{n_i(n_i-1)}{2} \int |\phi_i(\r)|^4 d\r\right\} \nonumber \\
 &+& \sum_{i<j}^{n_{orb}} 2 g n_i n_j \int |\phi_i(\r)|^2 |\phi_j(\r)|^2 d\r . \
\eeqn
The $N$-boson energy functional $E_N$ depends on the number $n_{orb}$ of orthogonal orbitals
in which the bosons reside,
the occupations $\{n_i\}$ of these orbitals
and the orbitals $\phi_i(\r)$ themselves which have to be determined {\it self-consistently}.
In order to find the ground state of the many-bosonic system
in the subspace of all possible configurations $\Psi$,  
one has to minimize the energy functional (\ref{energy_functional}) with respect to its arguments.
In practice, this results in $n_{orb}$ coupled, non-linear equations
which have to be solved {\it self-consistently}.
The occupations $\{n_i\}$ are determined variationally to minimize the energy.

The multi-orbital energy functional (\ref{energy_functional})
has been successfully employed and led 
to the prediction of (macroscopic) fragmentation of the ground state of repulsive condensates \cite{ALN_PRA}
and of additional quantum phases of cold bosonic atoms in optical lattices \cite{OAL_lattice}. 
In macroscopic fragmentation, a large number of atoms
reside in a small number of orbitals.
Specifically, three orbitals were 
considered in \cite{ALN_PRA} within Eq.~(\ref{energy_functional}).
We shall see below that, on the pathway from condensation to fermionization
the number of orbitals needed to describe the ground state gradually grows. 

In what follows, we study the ``evolution'' of a trapped bosonic system
as the interaction strength increases.
This will allow us to explore the pathway to fermionization in detail.
We will see that {\it in between} condensation and fermionization the {\it shape} of the 
confining potential is particularly important.
To demonstrate and explore the reachness of possibilities for a bosonic system before it fermionizes,
we chose as an illustrative 1D example the potential profile $V(x)$ shown in Fig.~1.
It is an asymmetric double-well potential with the left well being deeper and narrower 
than the right one. 
The problem is rescaled to the dimensionless coordinate $x$
in which the 1D many-body Hamiltonian is written as
$\hat H = \sum_i^N \left[- d^2/dx_i^2 + V(x_i)\right] + \sum_{i<j}^N \lambda_0 \delta(x_i-x_j)$. 

For transparency of presentation, 
in the numerical example presented we follow the pathway 
from condensation to fermionization of $N=12$ bosons.
We have obtained similar results for various numbers of bosons.
We begin with a small value of the inter-particle interaction $\lambda_0$.
For convenience we introduce the quantity $\lambda=\frac{\lambda_0(N-1)}{2\pi}$.
For $\lambda=1$, the energy functional (\ref{energy_functional})
is minimal when all atoms reside in one orbital,
thus recovering Gross-Pitaevskii theory. 
The bosonic system is a condensate which resides in the left, deeper potential well, see Fig.~1(a).
If we enforce the bosons to occupy more orbitals, 
the energy is found to increase.
In other words, for $\lambda=1$ the best mean-field is Gross-Pitaevskii theory.
It is instructive to briefly analyze the energy functional $E_N$, Eq.~(\ref{energy_functional}), 
for a general system of $N$ weakly-interacting trapped bosons.
For weakly-interacting bosons, the one-particle terms dominate $E_N$.
Hence, bosons prefer to accumulate in one orbital which is lowest in energy. 
This is exactly the Gross-Pitaevskii energy and   
the energy functional (\ref{energy_functional}) boils down 
to it in case of weakly-interacting bosons.

Next, when the inter-particle interaction is increased beyond $\lambda \approx 18$
our system becomes two-fold fragmented.
We remind that the fragmentation of a repulsive condensate in the 
ground state was discovered only recently \cite{ALN_PRA}, and 
is by itself a basic phenomena to be further explored.
For $\lambda=20$, ten bosons occupy one orbital which is localized in the left,
deeper well, whereas 2 bosons occupy now an orthogonal orbital localized in the right, 
shallower well, see Fig.~1(b).
For $\lambda=30$ the system remains two-fold fragmented with
8 bosons in the left well (orbital) and 4 bosons in right well (orbital).
For $\lambda=40$ and $50$ the two-fold fragmented ground state consists of
7 bosons in the left well and 5 bosons in the right well, see Fig.~1(c).
So far, in the regime up to $\lambda=50$ the system is a two-fold fragmented condensate. 
Each of the two orbitals occupied by the bosons is mainly located in one of 
the two distinct wells of the potential. 
If we enforce the bosonic system to occupy more orbitals, the energy becomes higher.
This means that the best mean-field here is a two orbital mean-field, which we denote as MF(2). 
In other words, $E_N$ obtains its minimum with two self-consistent orbitals and 
their corresponding occupations.

Let us briefly explain why fragmentation of the BEC occurs in the double-well potential.
The mean-field boson-repulsion term in the Gross-Pitaevskii energy functional [Eq.~(\ref{energy_functional}) with $n_{orb}=1$] 
scales with the number of particles as $\lambda_0 \cdot N(N-1)/2$.
When the repulsion between the bosons increases,
the energy quickly grows. 
If, however, we distribute the bosons between two orbitals localized in the left and right wells
with occupations of $n_1$ and $n_2=N-n_1$,
the main contribution from mean-field terms in (\ref{energy_functional}) 
scales as $\lambda_0 \cdot \left[ N(N-1)/2 - a n_1(N-n_1) \right]$, $a \sim 1$.
From a certain value of the interaction strength on, 
it becomes preferable to occupy a second orbital 
since the cost in kinetic energy is compensated by the smaller boson-repulsion energy.

What happens to the bosonic system when $\lambda$ is further increased? 
With increasing interaction, the distribution of the bosons between
two fragments (orbitals) becomes energetically unfavorable, 
and a third orbital starts to be occupied, see Fig.~1(d) for $\lambda=60$.
There are still 5 and 7 bosons in the left and right wells, respectively,
as found in the bifragmented condensate for $\lambda=40$ shown in Fig.~1(c),
but the bosons in the right well occupy now two orbitals with the occupations 6 and 1.
When $\lambda$ is further increased, the fragment in the right well further
breaks apart, whereas the fragment in the left well remains unchanged, see Fig.~1(e,f). 
For $\lambda=100$, almost all bosons in the right well reside in
{\it different} orbitals, whereas all bosons in the left well still reside in a single orbital.
This {\it ground state} of the system is very surprising.
It shows the {\it coexistence} of a coherent part in the left
well and a (almost) fully fermionized part in the right well.

We can explain this intriguing situation 
that the trapped boson system is found in a ground state
with coexisting condensed and fragmented parts as described above, see Fig.~1(f).
For this, we remind that the two fragments of the double-well system
are nearly non-overlapping and hence can be crudely viewed as two 
``almost independent'' condensates.
The effective strength $\gamma$ (see below) of the inter-particle interaction in the
right well is larger than that in the left well,
because the {\it density} in the right well is smaller.
Hence, the right fragment almost finishes ``fermionizing'' while the fragment in
the left well still remains ``condensed''. 

For $\lambda=110$, each boson in the right well occupies a different, orthogonal orbital. 
On the other end, the fragment in the left well just begins to spilt, with 4 and 1
bosons residing in two different orbitals, see Fig.~1(g).
For $\lambda=120$, one boson moves from the left to the right well
which now houses 8 bosons in 8 different orbitals,
and in the left well there is a 3-boson pack and 
a single additional boson, see Fig.~1(h). 
For $\lambda=150$, we find in the left well a 2-boson pack.
All remaining bosons, 2 more in the left well and 8 in the right well,
reside in different orbitals, see Fig.~1(i).
Finally, from $\lambda \approx 155$ on,
all bosons reside in different orthogonal orbitals --
the bosonic system has become fully {\it fermionized}, see Fig.~1(j).
It is instructive to express this critical value of $\lambda$
in terms of the dimensionless parameter $\gamma$ often employed in 1D \cite{Petrov} 
which reads here $\gamma=\lambda_0/2 n$, where $n$ is the boson density.
When this is done, we obtain for the {\it present} bosonic system
that onset of fermionization is at $\gamma \sim 9$,
a value that can be obtained nowadays experimentally 
(see Ref.~\cite{IB_nature}).

When more particles are trapped in the double-well trap employed here,
the physical picture obtained above remains essentially the same --  we checked it numerically.
The pathway we obtained starts, of course, from condensation,
passes through fragmentation, gradual build-up of fermionization and
eventually arrives at (full) fermionization.
The regime of coexistence of fully condensed and fermionized parts 
is relatively sensitive to the number of trapped bosons 
and also to the shape of the trap.
We note in this respect that a coexistence of $n_1$ condensed bosons
and $N-n_1$ fermionized bosons can be realized
by appropriately designing double-well traps for any numbers $N$ and $n_1<N$ of atoms. 

The description of a bosonic system as gradually occupying 
more orbitals with increasing scattering length,
all the way from ideal condensation ($n_{orb}=1$) to fermionization ($n_{orb}=N$) 
is physically very appealing.
It allows us to microscopically monitor the competition between 
one-body kinetic and potential energies and the two-body repulsion in a {\it general} trap potential.
What is important to discuss is how this pathway may be observed experimentally.
Shown in Fig.~2 is the density profile of the bosonic system,
$\rho(x)= \frac{1}{N} \sum_i^{n_{orb}} n_i |\phi_i(x)|^2$ which can be measured.
As the interaction strength increases and more orbitals becomes occupied,
 the density {\it gradually acquires more oscillations in real space}. 
The number of oscillations increases with increasing $\lambda$, see Fig.~2, 
until it saturates at $N$ -- the number of bosons in the system. 
This, as just stated above, is when the system becomes fermionized.
Another possibility to follow experimentally the pathway to fermionization could be 
to release the trap and register the interference
patters obtained for different $\lambda$'s 
which should gradually loose contrast as $\lambda$ grows.
  
The pathway from condensation to fermionization 
can be obtained for {\it any} bosonic system by minimizing the energy
functional (\ref{energy_functional}).  
Let us examine the pathway in terms of the energy functional $E_N$.
As the scattering length increases,
the two-body, mean-field energy terms start to dominate $E_N$
and the occupations $\{n_i\}$ of the bosons
would like to become smaller
in order to minimize the energy.
This leads to the occupation of more and more orbitals in the ground state.
Eventually, all $n_i=1$ and the last term 
in the braces of Eq.~(\ref{energy_functional}) becomes identically zero,
thus reversing the situation described by Gross-Pitaevskii theory 
where all bosons reside in a single orbital.
When this happens, i.e., when each boson resides in a different, orthogonal orbital,
the system becomes {\it fermionized}. 
What happens when the inter-particle interaction is further increased?
As this happens, the $n_{orb}=N$ occupied orbitals become more localized in
the sense that their overlaps 
(the orbitals seen in Fig.~1 overlap and hence exhibit negative values as well)
 are gradually reduced.
This prevents the last term in (\ref{energy_functional}) from diverging. 
At the extreme case of infinitely-strong interaction
there is no overlap at all between the orbitals, 
the last term in (\ref{energy_functional}) drops out
and the energy saturates.
This generic ``evolution'' from weakly- to strongly-interacting bosons holds true 
for {\it any trap potential and any dimension}, what makes the energy 
functional (\ref{energy_functional}) promising also beyond 1D.

In conclusion, in this work
the evolution of the density profile of a cold bosonic system
with increasing scattering length is followed.
With increasing scattering length the density profile
acquires more and more oscillations, 
until their number eventually equals to $N$, the number of bosons.
The ground state and density profile of a bosonic system strongly depend on the shape of the trap potential.
This allows one to design systems with an intriguing ground state,
e.g., for which one part is condensed and the other one is fermionized.
The employment of a more general {\it self-consistent mean-field theory} has been shown to 
be able to follow -- in real space -- the ``process'' of fermionization 
of a general cold bosonic system.
As to experimental realizations:  
Double-well potentials can be prepared by optical dipole traps
\cite{Kettele_disstilation} or by combining an optical lattice and strong harmonic confinement \cite{markus_DW}. 
The scattering length can be increased then by Feshbach-resonance techniques \cite{Feshbach_resonance}.
Alternatively, micorfabricated traps -- so-called atomchips -- offer 
a promising venue to more general low-dimensional trap potentials along with controllability of the
effective scattering length over a wide range of interaction strengths \cite{Jorg_Ron,Harvard_group}.

\acknowledgments

\noindent
We thank Alexej Streltsov for stimulating discussions.


\begin{figure}[ht] 
\includegraphics[width=10cm,angle=-0]{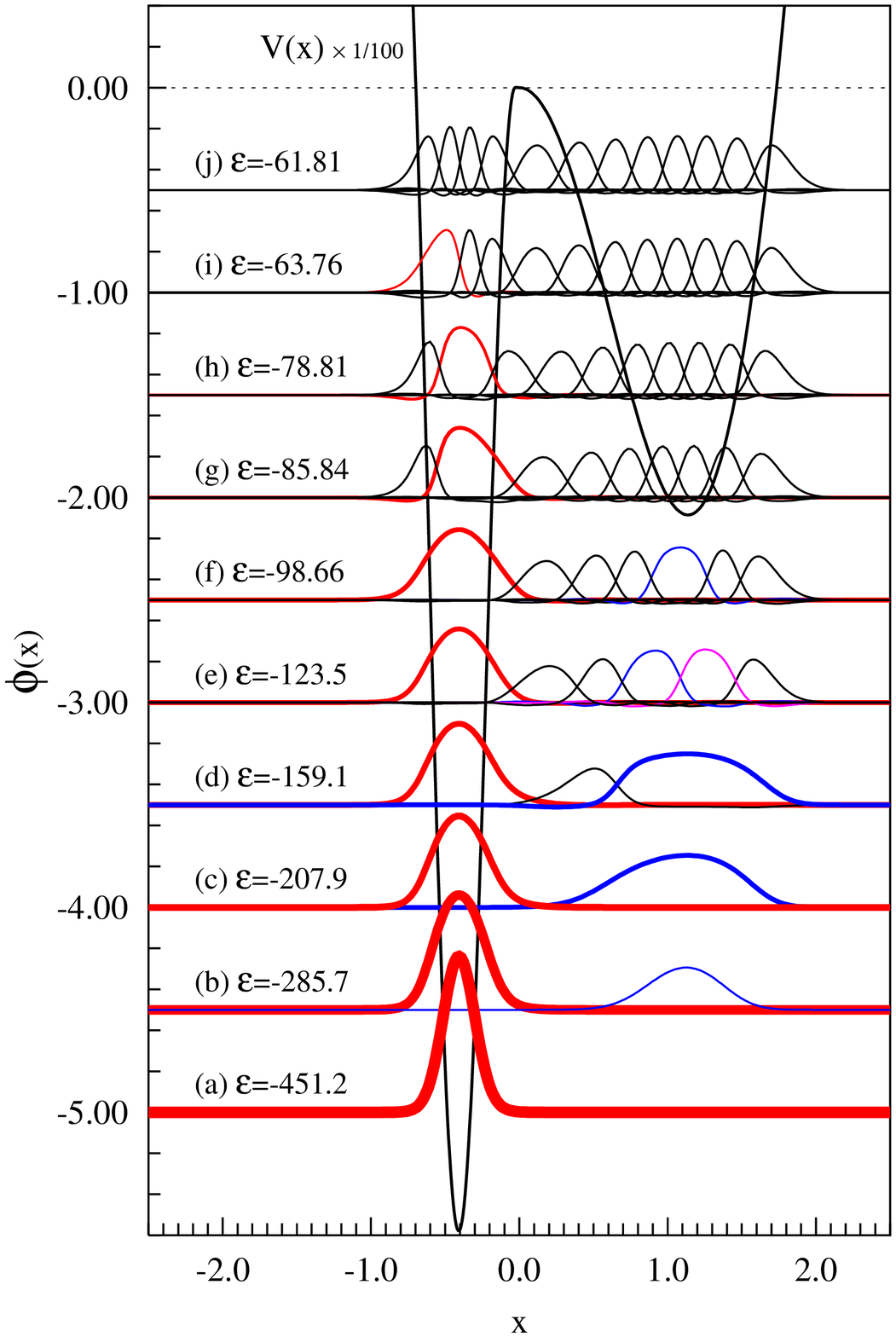}
\caption [kdv]{Ground-state wavefunction (orbitals $\phi_i(x)$ and occupations) 
of $N=12$ bosons in an asymmetric double-well potential $V(x)$
(scaled here by $1/100$) for interaction strengths:
(a) $\lambda=1$; 
(b) $\lambda=20$; 
(c) $\lambda=40$; 
(d) $\lambda=60$; 
(e) $\lambda=80$; 
(f) $\lambda=100$; 
(g) $\lambda=110$; 
(h) $\lambda=120$; 
(i) $\lambda=150$; 
(j) $\lambda=160$. 
As $\lambda$ increases, additional orbitals are needed to describe the ground state
which turns from a condensate to a fragmented condensate and eventually to a fermionized state.
For convenience, the $\sqrt{n_i/N} \phi_i(x)$ (scaled here by $1/3$ and shifted vertically and equidistantly) are plotted.
The energy-per-particle  $\varepsilon$ for each $\lambda$ is indicated on the plot.
}
\end{figure}


\begin{figure}[ht] 
\includegraphics[width=10cm,angle=-0]{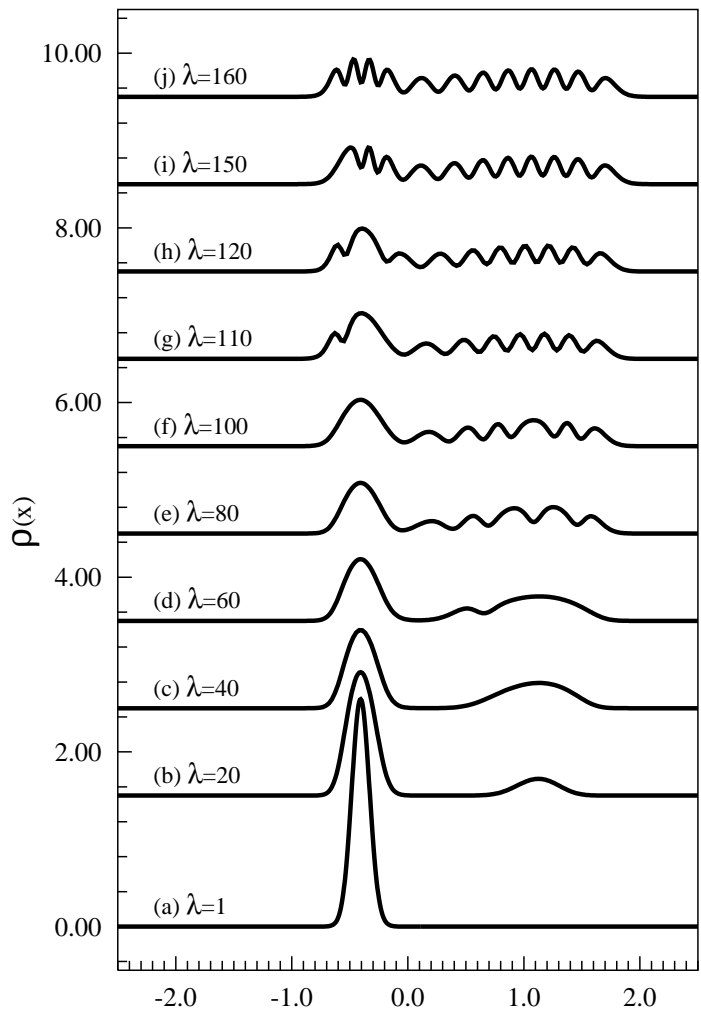}
\caption [kdv] {Ground-state density 
$\rho(x)= \frac{1}{N} \sum_i^{n_{orb}} n_i |\phi_i(x)|^2$
 of $N=12$ bosons in the asymmetric double-well potential $V(x)$ shown in Fig.~1
for interaction strengths in the range $\lambda=1$ to $160$.
For convenience, $\rho(x)$ is scaled here by $1/2$ and shifted vertically.}
\end{figure}

\ed